# Simultaneous Measurement of Thermal Conductivity, Heat Capacity, and Interfacial Thermal Conductance by Leveraging Negative Delay-Time Data in Time-Domain Thermoreflectance

Mingzhen Zhang[a], Tao Chen[a], Ao Zeng[a], Jialin Tang[b,c], Ruiqiang Guo[b], Puqing Jiang[a,*]

[a]*School of Power and Energy Engineering, Huazhong University of Science and Technology, Wuhan, Hubei 430074, China*

[b]*Thermal Science Research Center, Shandong Institute of Advanced Technology, Jinan, Shandong 250103, China*

[c]*Institute of Advanced Technology, Shandong University, Jinan, Shandong 250061, China*

**ABSTRACT**

Time-domain thermoreflectance (TDTR) is a widely used technique for characterizing the thermal properties of bulk and thin-film materials. Traditional TDTR analyses typically focus on positive delay time data for fitting, often requiring multiple-frequency measurements to simultaneously determine thermal conductivity and heat capacity. However, this multiple-frequency approach is cumbersome and may introduce inaccuracies due to inconsistencies across different frequency measurements. In this study, we propose a novel solution to these challenges by harnessing the often-overlooked negative delay time data in TDTR. By integrating these data points, we offer a streamlined, single-frequency method that simultaneously measures thermal conductivity, heat capacity, and interface thermal conductance for both bulk and thin-film materials, enhancing measurement efficiency and accuracy. We demonstrate the effectiveness of this method by measuring several bulk samples including sapphire, silicon, diamond, and $Si_{0.992}Ge_{0.008}$, and several thin-film samples including a 1.76-$\mu$m-thick gallium nitride (GaN) film epitaxially grown on a silicon substrate, a 320-nm-thick gallium oxide ($\varepsilon$-$Ga_2O_3$) film epitaxially grown on a silicon carbide substrate, and a 330-nm-thick tantalum nitride (TaN) film deposited on a sapphire substrate, all coated with an aluminum (Al) transducer layer on the surface. Our results show that the new method accurately determines the thermal conductivity and heat capacity of these samples as well as the Al/sample interface thermal conductance using a single modulation frequency, except for the $Si_{0.992}Ge_{0.008}$ sample. This study sheds light on the untapped potential of TDTR, offering a new, efficient, and accurate avenue for thermal analysis in material science.

**1. Introduction**

Thermal conductivity ($k$), heat capacity ($C$), and interfacial thermal conductance ($G$) are fundamental properties with wide-ranging implications in various scientific and technological fields. Thermal conductivity indicates how well a material conducts heat, offering valuable insights into its heat conduction efficiency. Heat capacity evaluates a material's capability to store thermal energy, and is a critical consideration in the design and enhancement of thermal storage systems [1]. Interface thermal conductance captures the intricacies of heat transfer dynamics at material boundaries, which is particularly important in the context of shrinking integrated circuits, where the interface density is escalating [2]. Accurate measurements of these properties are indispensable to aid researchers in precisely identifying bottlenecks, optimizing material designs, and advancing technologies for diverse scientific and engineering applications [3].

Time-domain thermoreflectance (TDTR) is a well-established and robust thermal measurement technique widely employed to measure the thermal properties of bulk and thin-film materials [4]. Typically, TDTR data processing focuses on phase signals within a positive delay period ranging from 0.1 - 4 ns (some systems extend this range to 8 ns) for fitting [5]. To simultaneously determine $k$ and $C$, dual-frequency measurements were employed [6, 7]. This approach capitalizes on the principle that when using a high modulation frequency of 10 MHz for the measurement, the heat transfer primarily exhibits one-dimensional characteristics in the cross-plane direction, with signals mainly sensitive to the material's cross-plane thermal effusivity ($\sqrt{k_z C}$). Conversely, at a low modulation frequency of 0.1 MHz, the heat transfer adopts a spherical cap-like behavior, making the signals sensitive to the material's in-plane thermal diffusivity ($k_r/C$) as well. For isotropic bulk materials, where $k$ equals both $k_r$ and $k_z$, dual-frequency measurements can effectively determine both $k$ and $C$. This approach has been proven effective in measuring bulk materials with a thermal diffusivity higher than $3.0 \times 10^{-6}$ m$^2$/s [6]. However, this method faces considerable challenges when using low-frequency TDTR measurements at 0.1 MHz. The primary issues arise from the considerable uncertainty in phase determination at these low frequencies, where pulse accumulation is significant [8]. Furthermore, the pronounced $1/f$ noise in the signals often leads to an unsatisfactory signal-to-noise ratio (SNR) for fitting.

Herein, we unveil the untapped potential of the often-overlooked negative delay time data in TDTR, demonstrating that its proper utilization enables simultaneous determination of $k$, $C$, and $G$ in a single-frequency measurement, thereby substantially enhancing measurement efficiency. To our knowledge, the existing literature only reports the use of negative delay time data in the beam-offset TDTR approach [9, 10], where the full-width-at-half-maximum (FWHM) of the out-of-phase ($V_{out}$) signal profile as a function of beam offset distance at a fixed negative delay time is used to extract the in-plane thermal diffusivity of the sample along the beam-offset direction. However, the practice of best-fitting phase signals as a function of delay time in the negative delay time range has not been previously reported.

In the subsequent sections, we first explain the principles of utilizing negative delay time data in TDTR to simultaneously determine $k$, $C$, and $G$ for bulk and thin-film materials. We then provide a detailed exploration of system noise and measurement uncertainty to ensure the reliability of our approach. To demonstrate this method, we applied it to measure several bulk samples including sapphire, silicon, diamond, and $Si_{0.992}Ge_{0.008}$, and several thin-film samples including a 1.76-$\mu$m-thick gallium nitride (GaN) film epitaxially grown on a silicon substrate, a 320-nm-thick gallium oxide ($\varepsilon$-$Ga_2O_3$) film epitaxially grown on a silicon carbide substrate, and a 330-nm-thick tantalum nitride (TaN) film deposited on a sapphire substrate, extracting the $k$ and $C$ of the samples as well as the $G$ of the metal transducer/sample interfaces. This study reveals the untapped potential of TDTR, opening a new avenue for efficient and accurate thermal analysis.

## 2. Methodologies

TDTR is a well-established technique with a history spanning over two decades. In-depth details of TDTR will not be provided here, but interested readers can find comprehensive information in various reliable reviews [4, 11] and classical research papers [12]. Here, we explain in detail the principles of simultaneously determining $k$, $C$, and $G$ from a single-frequency TDTR measurement of bulk and thin-film materials.

*2.1 Basic principles*

In TDTR experiments, when measuring a bulk material with a carefully chosen laser spot size and modulation frequency, the heat transfer regime is one-dimensional in the cross-plane direction with short delay times (~0.1 ns) and transitions to three-dimensional as the delay time extends to 12.5 ns. At 0.1 ns, the phase signal is sensitive only to the material's cross-plane thermal effusivity $\sqrt{k_z C}$. By 12.5 ns, the sensitivity for the in-plane thermal diffusivity $k_r/C$ increases. Therefore, by fitting the phase signals across the entire delay time range (0.1 to 12.5 ns), both $\sqrt{k_z C}$ and $k_r/C$ of the sample can be determined. For isotropic materials with $k = k_r = k_z$, both $k$ and $C$ can be derived from measurements. However, a standard TDTR setup typically captures data within a $t_d$ range of 0.1 to 4 ns. Although more sophisticated systems can extend this range to 8 ns using a pair of beam expanders, pushing it further to 12.5 ns requires a more intricate mechanical delay stage, introducing complexity and an increased risk of optical system errors.

Considering the periodic nature of TDTR signals, as illustrated in Fig. 1, with a period of 12.5 ns (corresponding to a laser pulse repetition rate of 80 MHz), the signal at a long delay time of 12.48 ns aligns with the signal at a negative delay time of -20 ps. Fitting the signal at -20 ps is therefore equivalent to fitting the signal at a long delay time of 12.48 ns. Hence, we propose that simultaneously fitting the phase signals from both the positive $t_d$ range (0.1~4 ns) and the negative $t_d$ range (-20 ps) should be sufficient to effectively decouple $k$ and $C$. When utilizing a high modulation frequency of ~10 MHz for the measurement, $G$ can also be effectively decoupled from $k$ and $C$ in the positive $t_d$ range of 0.1~4 ns, enabling simultaneous determination of these parameters in a single-frequency TDTR measurement.

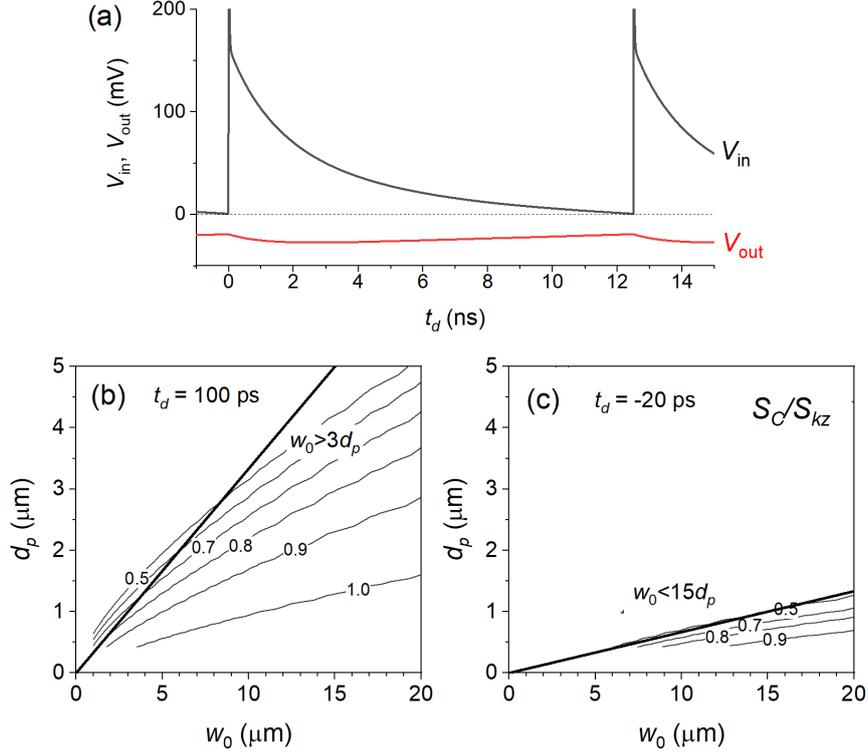

**Fig. 1.** (a) The periodic TDTR signal shows that the signal with a negative delay time is equivalent to the signal with a very long delay time. (b-c) The ratio of sensitivity coefficients of the TDTR phase signal to $C$ and $k_z$ of Si at two specific delay times of (b) $t_d = 0.1$ ns and (c) $t_d = -20$ ps, as a function of laser spot size $w_0$ and thermal penetration depth $d_p$ for the 100 nm Al/Si sample.

To simultaneously determine $k$ and $C$ of bulk material from a single-frequency measurement, it is crucial to design the experiment such that the heat transfer regime is primarily one-dimensional at 0.1 ns and transitions to three-dimensional at 12.5 ns. The key question is: what spot size and modulation frequency should be selected to achieve this transition in the heat-transfer regime?

To address this inquiry, Fig. 1(b-c) presents the ratio of the sensitivity coefficients of $C$ and $k_z$, denoted as $S_C/S_{k_z}$, for a 100 nm Al/Si sample. This sensitivity ratio is plotted against the laser spot radius ($w_0$) and thermal penetration depth ($d_p$) at fixed delay times of 0.1 ns and $-20$ ps, respectively. The sensitivity coefficient $S_\xi$ is defined as $S_\xi = \partial \ln R / \partial \ln \xi$, where $R = -V_{\text{in}}/V_{\text{out}}$ is the ratio signal, and $\xi$ represents any parameter under analysis. The thermal penetration depth $d_p$ is defined as $d_p = \sqrt{k/\pi f C}$, where $k$ and $C$ are the thermal properties of the substrate. Our focus was on the region where $S_C/S_{k_z} = 1 \pm 0.5$, indicating that the TDTR phase signal was almost equally sensitive to $k_z$ and $C$. This range ensures that the heat transfer regime is nearly one-dimensional in the through-plane direction.

Figure 1(b, c) demonstrates that using a laser spot size much larger than the thermal penetration depth is more favorable for establishing a one-dimensional cross-plane heat transfer regime. At $t_d = 0.1$ ns, the laser spot size $w_0$ needs to be larger than $3d_p$ to establish the one-dimensional heat transfer regime, whereas at $t_d = -20$ ps, achieving the same heat transfer regime requires a laser spot size larger than $15d_p$. When the laser spot size falls within the intermediate range ($3d_p \leq w_0 \leq 15d_p$), the heat transfer regime is one-dimensional at $t_d = 0.1$ ns but not at $-20$ ps. In such cases, the sensitivity to $k$ and $C$ can be decoupled by simultaneously fitting the phase signals from both positive and negative delay periods.

Sensitivity analysis was also conducted on various bulk samples with differing properties, confirming that this criterion is generally effective for high-conductivity materials with $k \geq 10$ W/(m·K). For substrates with low thermal conductivity measured within the range of $3d_p \leq w_0 \leq 15d_p$, signals in the negative delay time range become sensitive to the in-plane thermal diffusivity of the metal transducer layer instead. This sensitivity arises due to the substantial thermal impedance imposed by the low-conductivity substrate.

The outlined criterion also defines the measurable thermal diffusivity range as $\left(\frac{w_0}{15}\right)^2 \pi f \leq k/C \leq \left(\frac{w_0}{3}\right)^2 \pi f$ for a fixed laser spot size and modulation frequency configuration. For instance, using a typical laser spot size of 10 μm and a modulation frequency of 10 MHz for the measurement, and assuming a typical heat capacity of 2 MJ/(m³·K), the measurable thermal conductivity range is 28 - 700 W/(m·K). Reducing the laser spot size by half can extend the lowest measurable thermal conductivity to 7 W/(m·K).

When dealing with a new material with unknown thermophysical properties, an initial estimation of its thermal diffusivity is required to estimate $d_p$. This estimation allows for the initial selection of the laser spot size and modulation frequency for the measurement. The thermal diffusivity can then be updated based on the results from the first round of measurements, and if necessary, the laser spot size and modulation frequency can be refined in subsequent rounds until optimized measurements with the lowest measurement uncertainty are achieved.

This method can also be extended to measure thin-film materials on substrates, although it is based on slightly different principles. When measuring thin-film materials, factors such as $\sqrt{k_{z,f} C_f}$, $k_{r,f}/C_f$, and the areal heat capacitance ($h_f C_f$) of the thin film influence the heat diffusion process, and the sensitivity coefficients for different parameters of the thin film are correlated as $S_{k_{z,f}} + S_{h_f} = S_{k_{r,f}} + S_{C_f}$, see **Supplementary Information Section S1** for a detailed derivation. Among the three combined parameters, $\sqrt{k_{z,f} C_f}$, $k_{r,f}/C_f$, and $h_f C_f$, the sensitivity to $\sqrt{k_{z,f} C_f}$ mainly stems from the $V_{\text{out}}$ signal and remains invariant with the delay time. The sensitivity to $k_{r,f}/C_f$ is usually negligibly small, especially when measured using a high frequency of 10 MHz. In contrast, the sensitivity to $h_f C_f$ increases as the delay time extends from 0.1 to 12.5 ns, as the heat diffuses deeper into the underlying layers. Therefore, by simultaneously fitting the phase signals from both the short and long delay periods, both $\sqrt{k_{z,f} C_f}$ and $h_f C_f$ of the thin film can be determined. With the predetermined film thickness $h_f$, both the cross-plane thermal conductivity $k_{z,f}$ and heat capacity $C_f$ of the film can be derived.

*2.2 Noise consideration*

A notable challenge in our method is the exceptionally weak magnitude of the ratio signal during the negative delay period, particularly at high modulation frequencies. These weak signals are highly susceptible to interference from coherent and environmental noise, which complicates accurate measurements. In addition, pump leakage into the detector can introduce undesired signals, posing another potential source of error.

To address these challenges, we implemented several strategies to enhance the signal accuracy. First, we eliminated the pump leak by utilizing an optical parametric oscillator (OPO) to shift the wavelengths of the pump and probe and strategically applied a sharp-edge short-pass filter before the detector to effectively block the reflected pump beam. In addition, we employed a double lock-in scheme, modulating the probe path at a low frequency of 200 Hz, and extracted the modulated outputs of the radio-frequency (RF) lock-in using two computer-based lock-in modules. This scheme successfully eliminated the coherent background signals. Furthermore, we meticulously adjusted the relative positions of various equipment components, including the driver of the electro-optic modulator, the function generator, the photodiode detector, and the RF lock-in amplifier. We also applied electromagnetic shielding to the data transmission cable to minimize coherent pickup. These strategic adjustments collectively mitigated the impact of noise and coherent pickup on the signals, resulting in a substantial improvement in both signal-to-noise ratio (SNR) and signal accuracy.

In practice, we ensure that the outputs of $V_{\text{in}}$ and $V_{\text{out}}$ at a negative delay time are negligibly small, less than 1% of the normal signal level, when the pump, probe path, or both are blocked. See **Supplementary Information Section S2** for the demonstration.

*2.3 Uncertainty analysis*

Our approach involves concurrent fitting of two sets of signals with distinct orders of magnitude to extract multiple unknown parameters. Each set of signals was fitted using the least-squares regression method. This process can be mathematically expressed as the minimization of the product of the root mean squared (RMS) differences between each set of experimental signals and the model predictions:

$$\psi(\boldsymbol{U}) = \prod_{j=1}^{M} \sqrt{\frac{1}{N_j} \sum_{i=1}^{N_j} \left( \frac{f_j(\boldsymbol{U}, \boldsymbol{P}, t_{d,i})}{E_j(t_{d,i})} - 1 \right)^2} \tag{1}$$

where $E_j(t_{d,i})$ represents the $j$-th set of experimental signals at delay times $t_{d,i}$, and $f_j$ corresponds to the values evaluated by the thermal model. The vector $\boldsymbol{U}$ is a $n_u \times 1$ column vector of the unknown properties $u_1, u_2, \ldots, u_{n_u}$, whereas $\boldsymbol{P}$ is an $n_p \times 1$ column vector of the control parameters $p_1, p_2, \ldots, p_{n_p}$.

At the best fit, the gradient of $\psi$ should be zero for each element in $\boldsymbol{U}$:

$$\sum_{j=1}^{M} \left( \frac{\prod_{k \neq j} \text{RMS}_k}{2 \text{RMS}_j} \right) \left( \frac{1}{N_j} \sum_{i=1}^{N_j} \frac{2 \left( f_j(\boldsymbol{U}^*, \boldsymbol{P}^*, t_{d,i}) - E_j(t_{d,i}) \right)}{E_j(t_{d,i})^2} \frac{\partial f_j(\boldsymbol{U}^*, \boldsymbol{P}^*, t_{d,i})}{\partial u_l} \right) = 0,$$

$$\text{for } l = 1, 2, \ldots, n_u \tag{2}$$

Here, $\boldsymbol{P}^*$ is a randomly chosen set of possible control parameters, given the uncertainties associated with these input parameters, and $\boldsymbol{U}^*$ is the corresponding set of fitting parameters that results in the best fit. The uncertainties of unknown parameters can be determined from the distribution of all possible $\boldsymbol{U}^*$.

The covariance matrix of $\boldsymbol{U}^*$ can be expressed explicitly after some analytical derivations, as detailed in **Supplementary Information Section S3**. The covariance matrix $\text{Var}[\boldsymbol{U}^*]$ takes the form:

$$\text{Var}[\boldsymbol{U}^*] = \begin{pmatrix} \sigma_{u_1}^2 & \text{cov}[u_1, u_2] & ... \\ \text{cov}[u_2, u_1] & \sigma_{u_2}^2 & ... \\ \vdots & \vdots & \ddots \end{pmatrix} \qquad (3)$$

Here, the elements on the principal diagonal, $\sigma_{u_1}, \sigma_{u_2}, ..., \sigma_{u_{n_u}}$, represent the variance of the unknown parameters $u_1, u_2, ..., u_{n_u}$, and the off-diagonal elements, $\text{cov}[u_i, u_j]$, denote the covariance of $u_i$ and $u_j$. A correlation coefficient between $u_i$ and $u_j$ can be defined as $\rho_{u_i u_j} = \text{cov}[u_i, u_j]/\sigma_{u_i}\sigma_{u_j}$. If $\rho_{u_i u_j} = 0$, it indicates that the variables $u_i$ and $u_j$ are completely independent of each other. Conversely, if $\rho_{u_i u_j} = \pm 1$, it indicates that the variables $u_i$ and $u_j$ are fully correlated.

## 3 Results and Discussion

To prove the effectiveness of this method, we applied it to several bulk samples including sapphire, silicon, diamond, and Si$_{0.992}$Ge$_{0.008}$, and several thin-film samples including a 1.76-μm-thick gallium nitride (GaN) film epitaxially grown on a silicon substrate, a 320-nm-thick gallium oxide (ε-Ga$_2$O$_3$) film epitaxially grown on a silicon carbide substrate, and a 330-nm-thick tantalum nitride (TaN) film deposited on a sapphire substrate, simultaneously extracting $k$, $C$, and $G$ for each sample from a single-frequency measurement. The Si$_{0.992}$Ge$_{0.008}$ alloy was chosen in addition to the other bulk samples because previous TDTR measurements of semiconductor alloys showed frequency dependence in the determined thermal conductivity [13].

For illustrative purposes, the measurement details of the bulk Si sample and TaN film sample are comprehensively presented and discussed in Sections 3.1 and 3.2, respectively. Measurements of the other samples are briefly summarized in Section 3.3. The measurement of the Si$_{0.992}$Ge$_{0.008}$ alloy, which differs from the others, is discussed separately in Section 3.4. Section 3.5 provides a summary of all the measurement results.

*3.1 Bulk silicon*

The bulk silicon sample was coated with an aluminum (Al) transducer layer for TDTR measurement. The actual Al thickness was determined as $70 \pm 4$ nm using the picosecond acoustic method. A modulation frequency of 10.6 MHz and a laser spot radius of 7.2 μm were used for the measurement. The measured ratio signals $R = -V_{\text{in}}/V_{\text{out}}$ for the $t_d$ period of 0.1 to 4 ns and -50 to -10 ps are presented in Fig. 2(a) and (b), respectively. The sensitivity coefficients of these signals to $k_r$, $k_z$, and $C$ of silicon, as well as $G$ of the Al/Si interface, are illustrated in Fig. 2(c) and (d) as functions of $t_d$.

Figures 2(c) and (d) reveal that in the positive $t_d$ range of 0.1 to 4 ns, the sensitivity to $k_z$ and $C$ of silicon coincides, while the sensitivity to $k_r$ is almost zero, indicating that the signals in this $t_d$ range are predominantly sensitive to $\sqrt{k_z C}$ of silicon. Conversely, in the negative $t_d$ range of -50 to -10 ps, the signals become highly sensitive to $k_r/C$ instead, resulting in a negative sensitivity to $C$. Therefore, both $\sqrt{k_z C}$ and $k_r/C$ of silicon can be simultaneously determined if both the data from the positive and negative $t_d$ ranges are best fitted. For isotropic materials, where $k$ equals both $k_r$ and $k_z$, the thermal conductivity $k$ and heat capacity $C$ of the sample can be derived.

Furthermore, the sensitivity to $G$ of the Al/Si interface changes its sign from positive to negative in the $t_d$ range from 0.1 ns to 4 ns, thereby mainly influencing the slope of the phase signal in this delay period. Consequently, $G$ can also be simultaneously determined by best-fitting this group of signals. By fitting the two sets of signals shown in Fig.2(a) and (b), we obtain the three fitted parameters as $G_{Al/Si} = 130 \pm 6.6 \text{ MW}/(m^2 \cdot K)$, $k_{Si} = 135 \pm 9.8 \text{ W}/(m \cdot K)$, and $C_{Si} = 1.61 \pm 0.11 \text{ J}/(cm^3 \cdot K)$, each with an uncertainty of 5.1%, 7.2%, and 6.8%, respectively. These results agree well with the reference values [14, 15].

These measurement uncertainties are illustrated in Fig. 2(e-g), where the ellipses indicate the 95% confidence intervals for the best-fit parameters [16, 17]. Mathematically, these ellipses can be calculated as

$$\left[\frac{(x-x_0)\cos\varphi + (y-y_0)\sin\varphi}{\sqrt{\frac{\sigma_X^2+\sigma_Y^2+(\sigma_X^2-\sigma_Y^2)\sec(2\varphi)}{2}}}\right]^2 + \left[\frac{(y-y_0)\cos\varphi - (x-x_0)\sin\varphi}{\sqrt{\frac{\sigma_X^2+\sigma_Y^2-(\sigma_X^2-\sigma_Y^2)\sec(2\varphi)}{2}}}\right]^2 = 1, \qquad (4)$$

where $\varphi = \frac{1}{2}\arctan\left(\frac{2\text{cov}(X,Y)}{\sigma_X^2-\sigma_Y^2}\right)$, with $\sigma_X$, $\sigma_Y$, and $\text{cov}(X,Y)$ obtained from the covariance matrix in Eq. (3).

Correlation coefficients between $G_{Al/Si}$ and $k_{Si}$, $G_{Al/Si}$ and $C_{Si}$, and $k_{Si}$ and $C_{Si}$ are calculated as 0.67, 0.7, and -0.06, respectively. These values deviate from $\pm 1$, indicating effective decoupling of these parameters by incorporating phase signals of the negative delay time for best fitting.

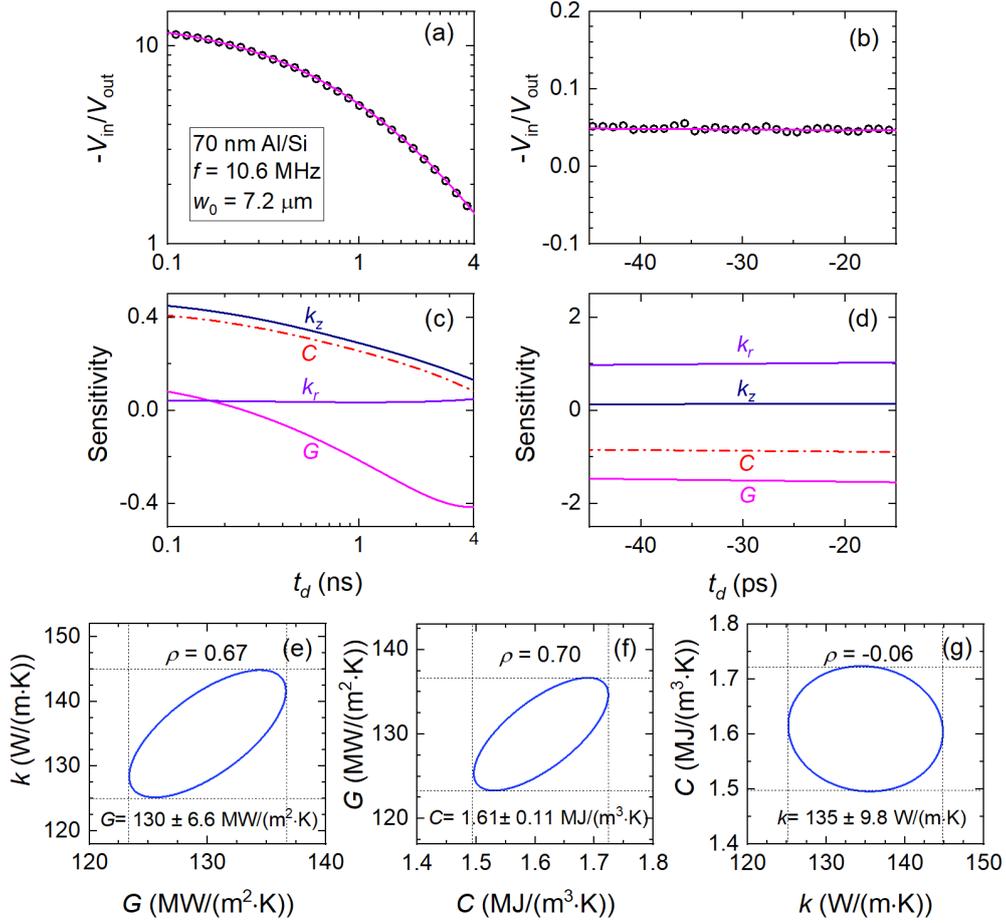

**Fig. 2.** (a, b) TDTR phase signals obtained from a 70 nm Al/Si sample, measured with a laser spot radius of 7.2 $\mu$m and a modulation frequency of 10.6 MHz, plotted against delay time. In (a), the delay time range spans from 0.1 to 4 ns, while in (b), it extends from -50 to -10 ps. The measured signals are represented by circle symbols, with calculated curves from the heat transfer model. (c, d) Illustration of the sensitivity coefficient variation of TDTR phase signals corresponding to (a, b) for each fitting parameter over delay time. (e-g) 95% confidence regions of $k$, $C$, and $G$ extracted by the simultaneous fitting of experimental signals in (a) and (b), presenting best-fitted values, uncertainties, and correlation coefficients $\rho$ of these three fitting parameters.

In conventional TDTR, the fitting process traditionally involves data exclusively in the positive delay time range, which allows for the determination of $k$ and $G$ provided that $C$ is a predefined input parameter, often resulting in an error greater than 10% for both $k$ and $G$. The new data processing method introduced here not only accommodates more fitting parameters but also produces more accurate results compared to the conventional TDTR data processing method.

*3.2 TaN film*

We further applied this method to simultaneously measure $k$ and $C$ of a tantalum nitride

(TaN) film sample. TaN films are widely used as sheet resistors in electronic and optical devices because of their near-zero temperature coefficient of resistance (TCR) [18]. Additionally, TaN has applications as a mechanical protective coating [19] and diffusion barrier [20]. In all these contexts, the thermal properties of TaN films are crucial for effective thermal management.

The TaN films in this study were prepared via radio-frequency (RF) reactive magnetron sputtering with a 3% $N_2$ flow ratio and deposited on a sapphire substrate. Previous studies indicated that TaN films produced using this method are predominantly in the $Ta_2N$ phase, offering stability against room-temperature oxidation [21].

For the TDTR measurements, a nominal 100 nm thick layer of Al was deposited on the TaN surface to serve as a transducer. Using the picosecond acoustic method [22], the actual thickness of the Al layer was determined to be 106 nm. The thickness of the TaN film was measured as 330 nm using cross-sectional imaging with scanning electron microscopy.

A modulation frequency of 5.2 MHz and a laser spot size of 7.5 $\mu$m were selected for the measurements. The signals measured at room temperature are shown in Fig. 3(a, b). The corresponding sensitivity coefficients of these signals to parameters including $k_{r,f}$, $k_{z,f}$, $C_f$, and $h_f$ of TaN, as well as the thermal boundary conductances ($G_1$ for the Al/TaN interface and $G_2$ for the TaN/sapphire interface), are depicted in Fig. 3(c) and (d) as functions of $t_d$.

Because of the high frequency of 5.2 MHz used, signals from both positive and negative $t_d$ ranges showed negligible sensitivity to $k_{r,f}$ and $G_2$, allowing these parameters to be excluded from further analysis. As shown in Fig. 3(c) and (d), the sensitivity coefficient curves for $k_{z,f}$, $C_f$, and $h_f$ are highly parallel, indicating a strong correlation among these parameters for each range of $t_d$. This parallelism suggests that fitting each range of data individually cannot decouple $k_{z,f}$ and $C_f$, thereby preventing their simultaneous determination. However, a critical observation is that while the sensitivity coefficient for $k_{z,f}$ remains almost constant, the sensitivity coefficient for $C_f$ changes its sign from positive to negative across the two ranges of $t_d$. Therefore, the simultaneous determination of $k_{z,f}$ and $C_f$ is possible by best fitting both data ranges.

In this set of measurements, limited attention was given to the negative delay time data, resulting in only a few data points in the range of -20 to -10 ps being available for analysis. This contrasts with our measurement of Si in Fig. 2(b), where many data points in the range of -50 to - 10 ps were collected to ensure a robust statistical average and minimize noise. However, as shown in Fig. 3(b), even the few data points in the range of -20 to -10 ps are sufficient to achieve a similar level of accuracy. The specific choice of these delay time ranges does not significantly impact the results due to the constancy of the signal in the range of -50 to -10 ps delay time.

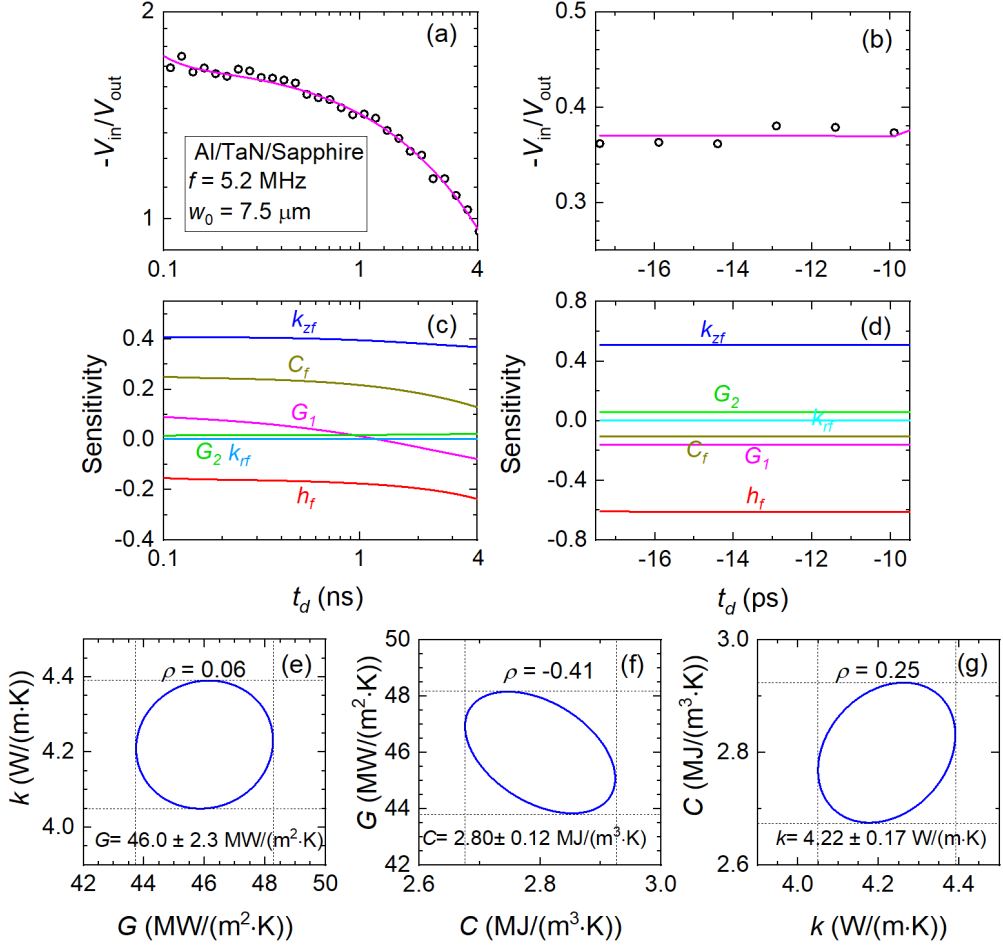

**Fig. 3.** (a, b) TDTR phase signals obtained from a 106 nm Al/330 nm TaN/Sapphire sample, measured with a laser spot radius of 7.5 $\mu$m and a modulation frequency of 5.2 MHz, plotted against delay time. In (a), the delay time range spans from 0.1 to 4 ns, while in (b), it covers the range from -20 to -10 ps. The measured signals are represented by circle symbols, with calculated curves from the heat transfer model. (c, d) Illustration of the sensitivity coefficient variation of TDTR phase signals corresponding to (a, b) for each fitting parameter over delay time. (e-g) 95% confidence regions of $k_{z,f}$, $C_f$, and $G_1$ extracted by the simultaneous fitting of experimental signals in (a) and (b), presenting best-fitted values, uncertainties, and correlation coefficients $\rho$ of these three fitting parameters.

Furthermore, the sensitivity to $G_1$ of the Al/TaN interface changes its sign from positive to negative in the $t_d$ range from 0.1 ns to 4 ns, thereby mainly influencing the slope of the phase signal in this delay period. Consequently, $G_1$ can be simultaneously determined by best fitting this group of signals. By fitting the two sets of signals shown in Fig. 3(a) and (b), we obtain the three fitted parameters as $G_{Al/TaN} = 46.0 \pm 2.3 \text{ MW}/(\text{m}^2 \cdot \text{K})$, $k_{TaN} = 4.22 \pm 0.17 \text{ W}/(\text{m} \cdot \text{K})$, and $C_{TaN} = 2.80 \pm 0.12 \text{ J}/(\text{cm}^3 \cdot \text{K})$, each with an uncertainty of 5.0%, 4.0%, and 4.4%, respectively. The small correlation coefficients among the three parameters of $G_{Al/TaN}$, $k_{TaN}$, and $C_{TaN}$, as illustrated in Fig. 3(e-g), suggests effective decoupling of these three parameters through the incorporation of the negative delay time data.

The same measurements were repeated at different temperatures from 80 to 400 K, with the measured temperature-dependent $k_{z,f}$, $C_f$, and $G_1$ values shown in Fig. 4.

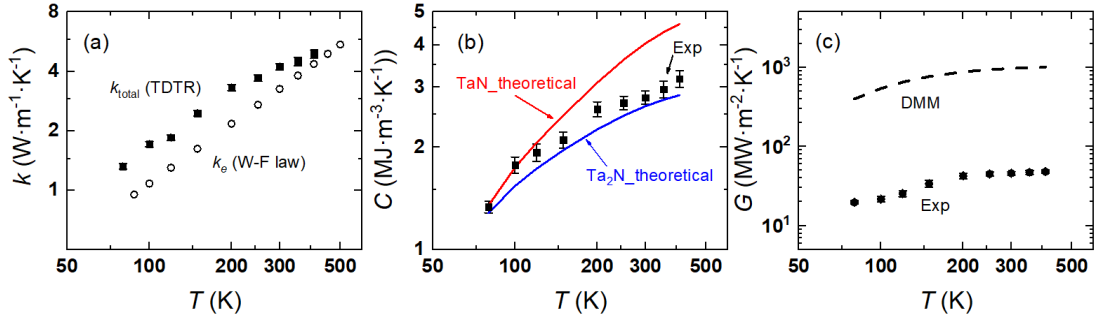

**Fig. 4.** (a) Total and electron contribution to the thermal conductivity of TaN film. (b) Volumetric heat capacity of TaN film from current experiments (symbol) and theoretical calculation (line). (c) Temperature-dependent thermal conductance of the Al/TaN interface.

The electronic contribution to the thermal conductivity of the TaN film ($k_e$) was estimated by measuring the temperature-dependent electrical resistance of the TaN film using the van der Paw method. This resistance was then converted to $k_e$ based on the Wiedemann-Franz law, assuming a theoretical Lorentz number value of $2.45 \times 10^{-8} \text{ W} \cdot \Omega/\text{K}^2$. The results are plotted as open symbols in Fig. 4(a) and compared with the total thermal conductivity measured using TDTR. The difference between these values represents the lattice contribution ($k_l$). The results indicate that the electronic contribution to the thermal conductivity increases with temperature, whereas the lattice contribution decreases. This is because, at higher temperatures, the number of thermally excited electrons near the Fermi level increases, contributing more to $k_e$. Simultaneously, the electron-phonon scattering process is enhanced, leading to a decreased $k_l$.

For additional reference purposes, we use the *ab initio* method to theoretically predict the heat capacity of a hexagonal ε-TaN crystal with the space group $P\bar{6}2m$ as $4.04 \text{ MJ}/(\text{m}^3 \cdot \text{K})$ and a hexagonal Ta$_2$N crystal with the space group $P\bar{3}m1$ as 2.64 MJ/(m$^3$·K). The measured heat capacity falls between these two theoretical values, as indicated in Fig. 4(b), which reinforces the consistency of our measurements because our sputtered TaN films are a mixture of polycrystalline and amorphous Ta-N phases dominated by the Ta$_2$N phase.

Figure 4(c) shows the temperature-dependent thermal conductance of the Al/TaN interface measured in this study, compared with theoretical predictions based on the Diffuse Mismatch Model (DMM) [23], which accounts for the full phonon dispersion of the materials but neglects the electron contribution to the interface thermal conductance. Both the experimental data and simulations indicate that the interface thermal conductance levels off as the temperature increases, although the simulations are approximately ten times higher in magnitude than the measurements. A possible reason for the measured values being significantly lower than the theoretical predictions could be the oxidation of the TaN films in air before Al deposition, resulting in an oxide interlayer between the Al and TaN.

*3.3 More measurement examples*

To further prove the effectiveness of the negative delay time TDTR method, we applied it to more bulk samples including sapphire and diamond, and more thin-film samples including a 1.76-$\mu$m-thick GaN film epitaxially grown on a silicon substrate, and a 320-nm-thick ε-Ga$_2$O$_3$ film epitaxially grown on a 4H silicon carbide substrate, with the measured results summarized below.

*3.3.1 Sapphire*

Figure 5(a1, b1) presents the TDTR signals of a sapphire sample coated with a 71-nm-thick Al transducer layer, measured using a spot size of 7.0 $\mu$m and a modulation frequency of 1.2 MHz. The sensitivity coefficients of these signals to $k_r$, $k_z$, and $C$ of sapphire, as well as $G$ of the Al/sapphire interface, are illustrated in Fig. 5(c1) and (d1) as functions of $t_d$.

Examination of the sensitivity coefficient curves reveals that, once again, signals in the positive $t_d$ range are highly sensitive to $k_z$ and $C$ of sapphire, both with positive coefficients. However, in the negative $t_d$ range, the sensitivity to $k_z$ diminishes to zero, leaving sensitivity only to $k_r$ and $C$ of sapphire with opposite signs. Therefore, incorporating the negative $t_d$ range data effectively decouples $k$ and $C$ of sapphire, enabling their simultaneous determination.

By fitting the two sets of signals shown in Fig. 5(a1) and (b1), we obtain $k_{Al_2O_3} = 37.4 \pm 3.1$ W/(m·K), $C_{Al_2O_3} = 3.08 \pm 0.23$ J/(cm$^3$·K), and $G_{Al/Al_2O_3} = 139 \pm 7$ MW/(m$^2$·K), each with an uncertainty of 8.3%, 7.5%, and 5%, respectively.

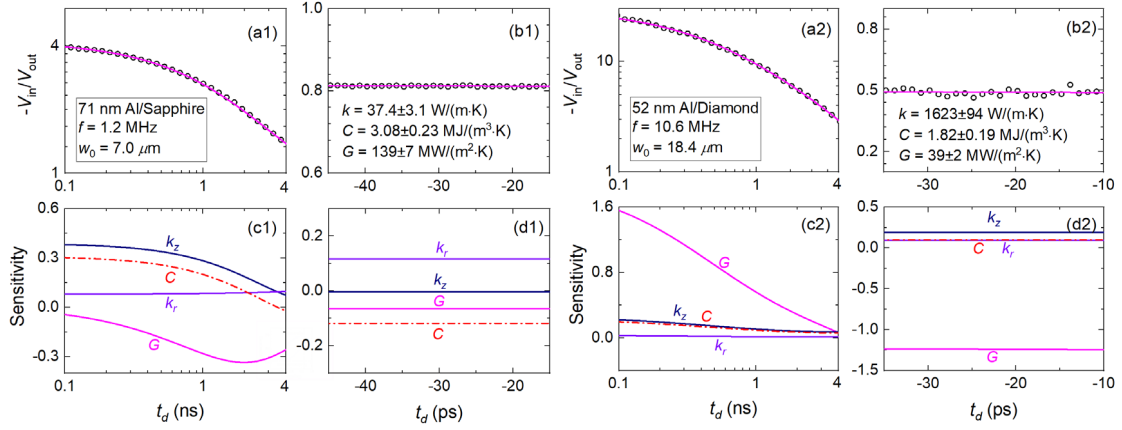

**Fig. 5.** (a1, b1) TDTR phase signals obtained from a 71 nm Al/sapphire sample, measured using a laser spot radius of 7.0 μm and a modulation frequency of 1.2 MHz, plotted against delay time. (a2, b2) TDTR phase signals obtained from a 52 nm Al/diamond sample, measured using a laser spot radius of 18.4 μm and a modulation frequency of 10.6 MHz, plotted against delay time. In (a1, a2), the delay time range spans from 0.1 to 4 ns, while in (b1, b2), it covers the negative range from -50 to -10 ps. The measured signals are represented by circle symbols, with calculated curves from the heat transfer model. (c1, c2, d1, d2) Illustration of the sensitivity coefficient variation of TDTR phase signals corresponding to (a1, a2, b1, b2) for each fitting parameter over delay time.

*3.3.2 Diamond*

Figure 5(a2, b2) displays the TDTR signals of a CVD diamond sample with a 52-nm-thick Al transducer layer, measured using an 18.4 μm spot size and a modulation frequency of 10.6 MHz. The sensitivity coefficients of these signals to $k_r$, $k_z$, and $C$ of diamond, as well as $G$ of the Al/diamond interface, are shown in Fig. 5(c2) and (d2) as functions of $t_d$.

Analysis of the sensitivity coefficient curves indicates that the signals exhibit high sensitivity to the Al/diamond interfacial thermal conductance $G$, with the sensitivity coefficient for $G$ transitioning sharply from a large positive value in the positive $t_d$ range to a large negative value in the negative $t_d$ range. Meanwhile, signals in the positive $t_d$ range are sensitive to $k_z$ and $C$ but not to $k_r$ of diamond, indicating sensitivity primarily to $\sqrt{k_z C}$ of diamond. In the negative $t_d$ range, the sensitivity to $C$ decreases while the sensitivity to $k_r$ increases, suggesting simultaneous sensitivity to $\sqrt{k_z C}$ and $k_r/C$ of diamond. Therefore, incorporating data from the negative $t_d$ range enables the simultaneous determination of $k$, $C$, and $G$ for this sample.

By fitting the two sets of signals shown in Fig. 5(a2) and (b2), we determine $k_{dia} = 1623 \pm 94$ W/(m·K), $C_{dia} = 1.82 \pm 0.19$ J/(cm$^3$·K), and $G_{Al/diamond} = 39 \pm 2$ MW/(m$^2$·K), with uncertainties of 5.8%, 10.4%, and 5%, respectively.

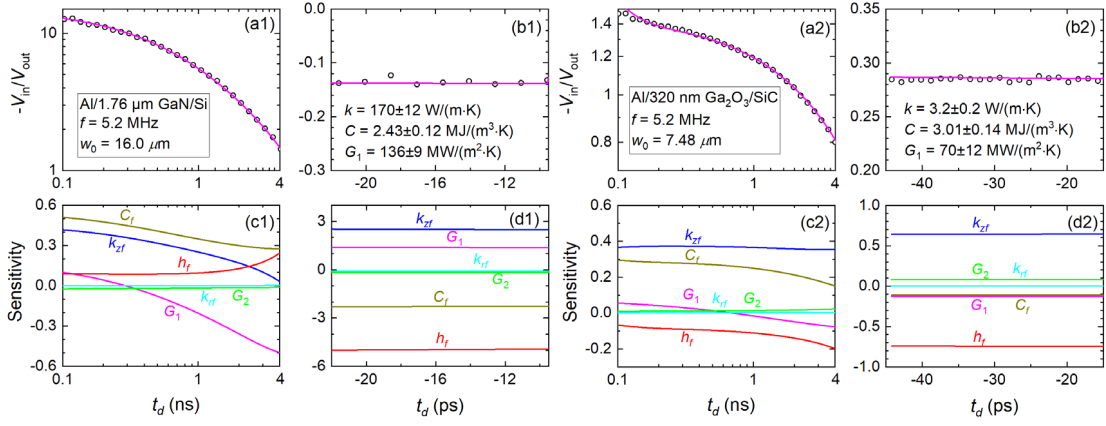

**Fig. 6.** (a1, b1) TDTR phase signals obtained from a 85 nm Al/1.76 $\mu$m GaN/Si sample, measured using a laser spot radius of 16.0 $\mu$m and a modulation frequency of 5.2 MHz, plotted against delay time. (a2, b2) TDTR phase signals obtained from a 121 nm Al/320 nm ε-Ga$_2$O$_3$/4H-SiC sample, measured using a laser spot radius of 7.48 $\mu$m and a modulation frequency of 5.2 MHz, plotted against delay time. In (a1, a2), the delay time range spans from 0.1 to 4 ns, while in (b1, b2), it covers the negative range from -50 to -10 ps. The measured signals are represented by circle symbols, with calculated curves from the heat transfer model. (c1, c2, d1, d2) Illustration of the sensitivity coefficient variation of TDTR phase signals corresponding to (a1, a2, b1, b2) for each fitting parameter over delay time.

*3.3.3 GaN film*

Figure 6(a1, b1) presents the TDTR signals of a 1.76-$\mu$m-thick GaN film, epitaxially grown on a Si substrate and coated with a 85-nm-thick Al transducer layer. These measurements were conducted using a 16.0 $\mu$m spot size and a modulation frequency of 5.2 MHz. The sensitivity coefficients of these signals to $k_r$, $k_z$, $h$, and $C$ of the GaN film, as well as $G_1$ of the Al/GaN interface and $G_2$ of the GaN/Si interface, are depicted in Fig. 6(c1) and (d1) as functions of $t_d$.

Analysis of the sensitivity coefficient curves indicates that the signals in both positive and negative $t_d$ ranges are insensitive to $k_r$ of the GaN film and $G_2$ of the GaN/Si interface, thereby excluding these parameters from the analysis. The sensitivity coefficient to $G_1$ of the Al/GaN interface transitions from positive at 0.1 ns to negative as $t_d$ increases beyond 0.3 ns, suggesting that $G_1$ mainly influences the signal gradient in the positive $t_d$ range of 0.1 to 1 ns. This distinct behavior allows for the determination of $G_1$ by fitting signals in this $t_d$ range.

In the positive $t_d$ range, the signals are sensitive to $k_z$ and $C$ of GaN, both with positive coefficients. In the negative $t_d$ range, the signals remain highly sensitive to $k_z$ and $C$ of GaN, but with opposite coefficients. Therefore, $k_z$ and $C$ of the GaN film can be determined by fitting data from both the positive and negative $t_d$ ranges.

By fitting the two sets of signals shown in Fig. 6(a1) and (b1), we determine $k_{z,\text{GaN}} = 170 \pm 12 \text{ W/(m} \cdot \text{K)}$, $C_{\text{GaN}} = 2.43 \pm 0.12 \text{ J/(cm}^3 \cdot \text{K)}$, and $G_{\text{Al/GaN}} = 136 \pm 9 \text{ MW/(m}^2 \cdot \text{K)}$, with uncertainties of 7%, 5%, and 6.6%, respectively.

*3.3.4 ε-Ga₂O₃ film*

Figure 6(a2) to (d2) presents another measurement example of a 320-nm-thick ε-Ga$_2$O$_3$ film, epitaxially grown on a 4H-SiC substrate and coated with a 121-nm-thick Al transducer layer. Figure 6(a2, b2) shows the measurement signals, which were obtained from TDTR measurements using a 7.48 $\mu$m spot size and a modulation frequency of 5.2 MHz. The sensitivity coefficients of these signals to $k_r$, $k_z$, $h$, and $C$ of the ε-Ga$_2$O$_3$ film, as well as $G_1$ of the Al/Ga$_2$O$_3$ interface and $G_2$ of the Ga$_2$O$_3$/SiC interface, are depicted in Fig. 6(c2) and (d2) as functions of $t_d$.

Analysis of the sensitivity coefficient curves indicates that the signals in both positive and negative $t_d$ ranges have negligible sensitivity to $k_r$ of the ε-Ga$_2$O$_3$ film and $G_2$ of the Ga$_2$O$_3$/SiC interface, thereby excluding these parameters from the analysis. The sensitivity coefficient to $G_1$ of the Al/Ga$_2$O$_3$ interface transitions from positive at 0.1 ns to negative as $t_d$ increases to 4 ns, suggesting that $G_1$ mainly influences the signal gradient in this $t_d$ range. This distinct behavior allows for the determination of $G_1$ by fitting signals in this $t_d$ range.

In the positive $t_d$ range, the signals are sensitive to $k_z$ and $C$ of Ga$_2$O$_3$, both with positive coefficients. In the negative $t_d$ range, the sensitivity to $k_z$ of Ga$_2$O$_3$ remains positive, but the sensitivity to $C$ changes to negative. Therefore, $k_z$ and $C$ of the Ga$_2$O$_3$ film can be determined by fitting data from both the positive and negative $t_d$ ranges.

By fitting the two sets of signals shown in Fig. 6(a2) and (b2), we determine $k_{z,\text{Ga}_2\text{O}_3} = 3.2 \pm 0.2 \text{ W/(m} \cdot \text{K)}$, $C_{\text{Ga}_2\text{O}_3} = 3.01 \pm 0.14 \text{ J/(cm}^3 \cdot \text{K)}$, and $G_{\text{Al/Ga}_2\text{O}_3} = 70 \pm 12 \text{ MW/(m}^2 \cdot \text{K)}$, with uncertainties of 6.3%, 5%, and 17%, respectively.

*3.4 Si$_{0.992}$Ge$_{0.008}$ alloy*

The $k$ and $C$ of the aforementioned samples can also be measured using the dual-frequency TDTR approach, provided no frequency-dependent systematic errors bias the results. However, this approach is unsuitable for materials exhibiting frequency-dependent thermal conductivity, such as semiconductor alloys [13]. Various theories, including non-equilibrium phonon transport [24] and Lévy superdiffusion dynamics [25], have been proposed to explain this phenomenon. In this study, we selected a Si$_{0.992}$Ge$_{0.008}$ alloy in addition to the other bulk samples to check the validity of the negative delay time TDTR method for simultaneous measurements of $k$ and $C$ from a single frequency measurement.

Figure 7 (a, b) presents the TDTR signals of a Si$_{0.992}$Ge$_{0.008}$ sample coated with a 105-nm-thick Al transducer layer, measured using a spot size of 8.0 $\mu$m and a modulation frequency of 9.8 MHz. The sensitivity coefficients of these signals to $k_r$, $k_z$, and $C$ of Si$_{0.992}$Ge$_{0.008}$, as well as $G$ of the Al/Si$_{0.992}$Ge$_{0.008}$ interface, are illustrated in Fig. 7(c) and (d) as functions of $t_d$.

Examining the sensitivity coefficient curves reveals that signals in the positive $t_d$ range are sensitive only to $\sqrt{k_z C}$ but not $k_r$ of the substrate. In the negative $t_d$ range, the sensitivity to $k_r$ of the substrate increases, causing the sensitivities to $k_z$ and $C$ of the substrate to deviate from each other. Therefore, incorporating the negative $t_d$ range data decouples $k$ and $C$ of Si$_{0.992}$Ge$_{0.008}$, enabling their simultaneous determination.

By best fitting the two sets of signals shown in Fig. 7(a) and (b) simultenously, we obtain $k_{\text{SiGe}} = 58.8 \pm 5 \text{ W/(m} \cdot \text{K)}$, $C_{\text{SiGe}} = 0.9 \pm 0.07 \text{ J/(cm}^3 \cdot \text{K)}$, and $G_{\text{Al/SiGe}} = 213 \pm 10 \text{ MW/}$(m$^2 \cdot$ K). The obtained $k$ and $C$ of Si$_{0.992}$Ge$_{0.008}$, however, deviate significantly from expectations. We estimate the heat capacity of the Si$_{0.992}$Ge$_{0.008}$ alloy to be $1.639 \text{ J/(cm}^3 \cdot \text{K)}$ based on the virtual crystal approximation, with the heat capacities of Si and Ge obtained from TPRC database [26]. Literature also reports a thermal conductivity of only $46 \text{ W/(m} \cdot \text{K)}$ for the Si$_{0.992}$Ge$_{0.008}$ alloy [27].

Assuming the heat capacity of $1.639 \text{ J/(cm}^3 \cdot \text{K)}$ is known for the Si$_{0.992}$Ge$_{0.008}$ alloy, best fitting the positive $t_d$ range data yields $k_{\text{SiGe}} = 33.5 \pm 3.3 \text{ W/(m} \cdot \text{K)}$ and $G_{\text{Al/SiGe}} = 200 \pm 10 \text{ MW/(m}^2 \cdot \text{K)}$. However, the best-fitted simulation curve cannot fit the signals in the negative $t_d$ range simultaneously, as shown by the blue dashed curves in Fig. 7 (a and b). This experiment suggests that the underlying mechanisms causing TDTR-frequency-dependent thermal conductivity measurement also causes problems in simultaneously fitting both the positive and negative $t_d$ range data. Forcing their simultaneous fitting would yield erroneous results.

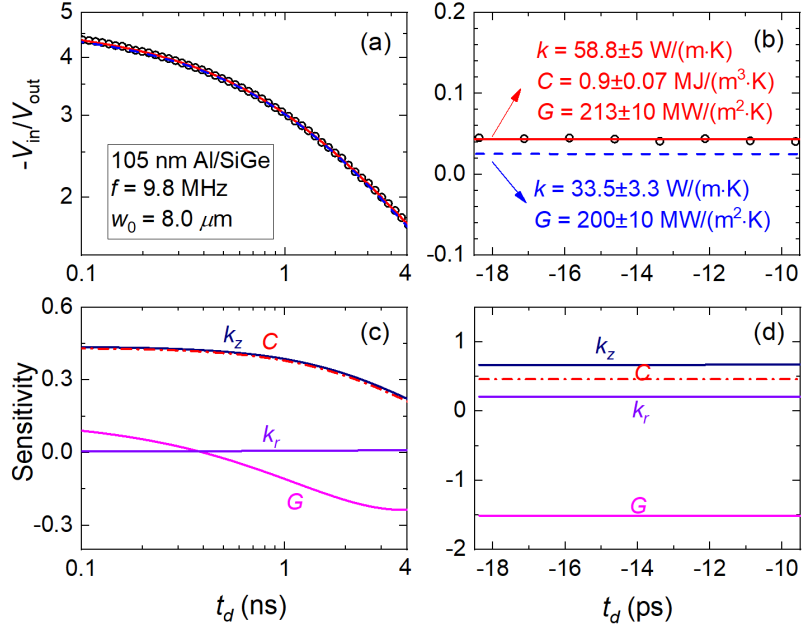

**Fig. 7.** (a, b1) TDTR phase signals obtained from a 105 nm Al/ $Si_{0.992}Ge_{0.008}$ sample, measured using a laser spot radius of 8.0 $\mu$m and a modulation frequency of 9.8 MHz, plotted against delay time. In (a), the delay time range spans from 0.1 to 4 ns, while in (b), it covers the negative range from -20 to -10 ps. The measured signals are represented by circle symbols. The red solid curves are obtained by best fitting both ranges of data, treating $k$, $C$, and $G$ as fitting parameters, while the blue dashed curves are obtained by best fitting only the positive delay time range data, with $k$ and $G$ as fitting parameters. (c, d) Illustration of the sensitivity coefficient variation of TDTR phase signals corresponding to (a, b) for each fitting parameter over delay time.

*3.5 Summary*

Table 1 summarizes the $k$ and $C$ of all the samples measured in this study, and compared with the available literature values. The results all compare well except for the $Si_{0.992}Ge_{0.008}$ alloy.

**Table I.** The literature and measured values for thermal conductivity ($k$) and volumetric heat capacity ($C$) of different reference samples.

| Sample | Literature | | Current | |
|---|---|---|---|---|
| | $C$ (MJ/(m³·K)) | $k$ (W/(m·K)) | $C$ (MJ/(m³·K)) | $k$ (W/(m·K)) |
| Sapphire | 3.05 ± 0.1 [28] | 35 ± 3.5 [29] | 3.08 ± 0.23 | 37.4 ± 3.1 |
| Silicon | 1.665 ± 0.05 [30] | 142 ± 2.8 [15] | 1.61 ± 0.11 | 135 ± 9.8 |
| Diamond | 1.93 ± 0.077 [31] | 1890 ± 173 [31] | 1.82 ± 0.19 | 1623 ± 94 |
| GaN | 2.6 ± 0.08 [32] | 185 ± 20 [33] | 2.43 ± 0.12 | 170 ± 12 |
| ε-Ga$_2$O$_3$ | 2.5~3.3 [34] | 3.15~4.39 [35] | 3.01 ± 0.14 | 3.2 ± 0.2 |
| TaN | 2.76 ± 0.31 [21] | 4.25 ± 0.33 [21] | 2.80 ± 0.12 | 4.22 ± 0.17 |
| Si$_{0.992}$Ge$_{0.008}$ | 1.637 ± 0.03 [26] | 46 [27] | 0.9 ± 0.07 | 58.8 ± 5 |

## 4. Conclusions

In this study, we introduced an innovative approach to TDTR that leverages negative delay time data to enhance the measurement of $k$, $C$, and $G$ in bulk and thin-film materials. By shifting from the conventional dual-frequency method to a streamlined single-frequency measurement, our approach addresses and overcomes the challenges associated with frequency inconsistencies and phase determination uncertainties at low frequencies. Our experimental validation on several bulk samples including sapphire, silicon, and diamond, and several thin film samples including a 330-nm-thick TaN film on a sapphire substrate, a 1.76-μm-thick GaN film on a Si substrate, and a 320-nm-thick ε-Ga$_2$O$_3$ film on a 4H-SiC substrate, demonstrated the efficacy of this method in accurately determining $k$, $C$, and $G$ for bulk and thin films. The application on a Si$_{0.992}$Ge$_{0.008}$ alloy, however, is unsuccessful, possibly due to the same underlying mechanisms that cause TDTR-frequency-dependent thermal conductivity in these materials. Nevertheless, this novel utilization of negative delay time data not only simplifies the TDTR measurement process but also enhances its accuracy and efficiency. These advancements have opened new avenues for thermal analysis in material science, facilitating more precise and reliable characterization of thermal properties in a wide range of materials. Future work will explore the application of this method to other material systems and further refine the technique to push the boundaries of the thermal property measurements.

**Supplementary Material**

Refer to the supplementary material for a detailed derivation of the relationship between the sensitivity coefficients of different parameters, a demonstration of noise reduction, and a comprehensive derivation of the formula for the uncertainty analysis.

**Acknowledgments**

The authors sincerely thank Prof. Ronggui Yang and Prof. Xin Qian for their valuable discussions and support with the equipment. We also express our gratitude to Prof. Junjun Wei from the Beijing University of Science and Technology for providing the TaN and GaN film samples tested in this study, and to Prof. Yanli Pei from Sun Yat-sen University for providing the GaO film sample. P.J. acknowledges support from the National Natural Science Foundation of China (NSFC) under Grant No. 52376058.

**Data Availability**

The data supporting the findings of this study are available from the corresponding authors upon reasonable request.

**Declaration of Competing Interest**

The authors have no conflicts to disclose.

**Author Contributions**

Mingzhen Zhang: Data curation (equal); Formal analysis (equal); Investigation (equal); Methodology (equal); Validation (equal); Writing – original draft (equal). Tao Chen: Formal analysis (equal); Investigation (equal). Ao Zeng: Data curation (equal); Investigation (equal). Ruiqiang Guo: Investigation (equal); Methodology (equal). Puqing Jiang: Conceptualization (lead); Formal analysis (equal); Methodology (equal); Validation (equal); Project administration (equal); Writing – original draft (equal); Writing – review & editing (lead).

# Supplementary Information

# Simultaneous Measurement of Thermal Conductivity, Heat Capacity, and Interfacial Thermal Conductance by Leveraging Negative Delay-Time Data in Time-Domain Thermoreflectance

Mingzhen Zhang[a], Tao Chen[a], Ao Zeng[a], Jialin Tang[b, c], Ruiqiang Guo[b], Puqing Jiang[a, *]

[a]*School of Power and Energy Engineering, Huazhong University of Science and Technology, Wuhan, Hubei 430074, China*
[b]*Thermal Science Research Center, Shandong Institute of Advanced Technology, Jinan, Shandong 250103, China*
[c]*Institute of Advanced Technology, Shandong University, Jinan, Shandong 250061, China*

**Section SI. Relationship Between Sensitivity Coefficients of Different Parameters**

*S1.1 Proof of key parameters $k_z C$, $k_r/C$, and $hC$*

We start with the thermal diffusion equation for each layer of the homogeneous and transversely isotropic media:

$$C\frac{\partial T}{\partial t} = \frac{k_r}{r}\frac{\partial}{\partial r}\left(r\frac{\partial T}{\partial r}\right) + k_z\frac{\partial^2 T}{\partial z^2} \quad (S1)$$

where $k_r$ and $k_z$ are the thermal conductivities of the sample in the radial and through-plane directions, respectively, and $C$ is the volumetric heat capacity. Applying the Fourier transform to the time variable $t$ and the Hankel transform to the radial coordinate $r$, this parabolic partial differential equation can be simplified into an ordinary differential equation as:

$$\frac{\partial^2 \Theta}{\partial z^2} = \lambda^2 \Theta \quad (S2)$$

where $\lambda = \sqrt{4\pi^2\rho^2 k_r/k_z + i\omega C/k_z}$, $\rho$ is the Hankel transform variable, and $\omega$ is the angular frequency. The general solution of Eq. (S2) can be written as:

$$\Theta = e^{\lambda z}B^+ + e^{-\lambda z}B^- \quad (S3)$$

where $B^+$ and $B^-$ are complex constants to be determined based on the boundary conditions.

The heat flux can be obtained from the temperature (Eq. (S3)) and Fourier's law of heat conduction $Q = -k_z(d\Theta/dz)$ as:

$$Q = \gamma(-e^{\lambda z}B^+ + e^{-\lambda z}B^-) \tag{S4}$$

where $\gamma = k_z \lambda = \sqrt{4\pi^2 \rho^2 (k_r/C) \cdot (k_z C) + i\omega k_z C}$.

Equations. (S3) and (S4) can be rewritten in matrix as:

$$\begin{bmatrix} B^+ \\ B^- \end{bmatrix}_i = \frac{1}{2\gamma_i} \begin{bmatrix} \gamma_i & -1 \\ \gamma_i & 1 \end{bmatrix} \begin{bmatrix} \Theta \\ Q \end{bmatrix}_{i, z=0}$$

$$\begin{bmatrix} \Theta \\ Q \end{bmatrix}_{i, z=h_i} = \begin{bmatrix} 1 & 1 \\ -\gamma_i & \gamma_i \end{bmatrix} \begin{bmatrix} e^{\lambda_i h_i} & 0 \\ 0 & e^{-\lambda_i h_i} \end{bmatrix} \begin{bmatrix} B^+ \\ B^- \end{bmatrix}_i \tag{S5}$$

here, $h_i$ represents the thickness of the *i*-th layer. It is clear from Eq. (S5) that elements determining the transfer matrix are $\gamma$ and $\lambda h$. Note that $\gamma = \sqrt{4\pi^2 \rho^2 (k_r/C) \cdot (k_z C) + i\omega(k_z C)}$, and $\lambda h = (hC)\sqrt{4\pi^2 \rho^2 (k_r/C)/(k_z C) + i\omega/(k_z C)}$. Therefore, properties of each layer that affect the heat diffusion across multilayers are the three combined parameters of $k_z C$, $k_r/C$, and $hC$.

*S1.2 Proof of the relationship $S_{k_{zf}} + S_{h_f} = S_{k_{rf}} + S_{C_f}$*

The sensitivity coefficients for each parameter are defined as the ratio of the percentage change in the experimental signal $R$ to the percentage change in the parameter $\xi$:

$$S_\xi = \frac{\partial \ln R}{\partial \ln \xi} = \frac{\xi}{R}\frac{\partial R}{\partial \xi} \tag{S6}$$

Since the experimental signal $R$ depends on the three combined parameters of $k_z C$, $k_r/C$, and $hC$, i.e., $R = f(k_{z,f}C_f, \frac{k_{r,f}}{C_f}, h_f C_f)$, according to the chain rule, there should be:

$$\frac{\partial R}{\partial k_{zf}} = C_f \frac{\partial R}{\partial (k_{zf}C_f)} \tag{S7}$$

$$\frac{\partial R}{\partial k_{rf}} = \frac{1}{C_f} \frac{\partial R}{\partial (k_{rf}/C_f)} \tag{S8}$$

$$\frac{\partial R}{\partial h_f} = C_f \frac{\partial R}{\partial (h_f C_f)} \tag{S9}$$

$$\frac{\partial R}{\partial C_f} = k_{zf}\frac{\partial R}{\partial(k_{zf}C_f)} - \frac{k_{rf}}{C_f^2}\frac{\partial R}{\partial(k_{rf}/C_f)} + h_f\frac{\partial R}{\partial(h_fC_f)} \tag{S10}$$

Substituting Eqs. (S7) to (S10) into the definitions of each sensitivity coefficient, we have:

$$S_{k_{zf}} = \frac{k_{zf}}{R}\frac{\partial R}{\partial k_{zf}} = \frac{k_{zf}C_f}{R}\frac{\partial R}{\partial(k_{zf}C_f)} \tag{S11}$$

$$S_{k_{rf}} = \frac{k_{rf}}{R}\frac{\partial R}{\partial k_{rf}} = \frac{k_{rf}/C_f}{R}\frac{\partial R}{\partial(k_{rf}/C_f)} \tag{S12}$$

$$S_{h_f} = \frac{h_f}{R}\frac{\partial R}{\partial h_f} = \frac{h_fC_f}{R}\frac{\partial R}{\partial(h_fC_f)} \tag{S13}$$

$$S_{C_f} = \frac{C_f}{R}\frac{\partial R}{\partial C_f} = \frac{k_{zf}C_f}{R}\frac{\partial R}{\partial(k_{zf}C_f)} - \frac{k_{rf}/C_f}{R}\frac{\partial R}{\partial(k_{rf}/C_f)} + \frac{h_fC_f}{R}\frac{\partial R}{\partial(h_fC_f)} \tag{S14}$$

Combining Eqs. (S11) to (S14), there is the following relationship:

$$S_{k_{zf}} + S_{h_f} = S_{k_{rf}} + S_{C_f} \tag{S15}$$

**Section S2. Uncertainty analysis of multi-signal fitting multi-parameters**

Figure S1 shows the $V_{in}$ and $V_{out}$ signals obtained during a 10.6 MHz measurement under different conditions: (1) blocking the pump path, (2) blocking the probe path, (3) blocking both the pump and probe paths, and (4) normal signal measurement. The results demonstrate that the picked-up noise is negligible compared to the normal signal level, even though the overall signal level is low.

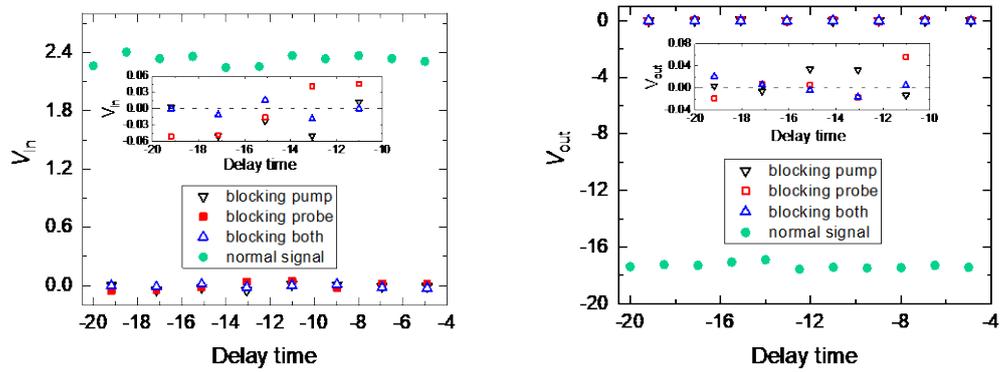

**Fig. S1.** The $V_{in}$ and $V_{out}$ signals obtained during a 10.6 MHz measurement under different conditions: (1) blocking the pump path, (2) blocking the probe path, (3) blocking both the pump and probe paths, and (4) normal signal measurement, to showcase the signal levels compared to the picked-up noises.

**Section S3. Uncertainty Analysis of Multi-Parameter Extraction from Multi-Signal Fitting**

In processing the data, we extract multiple parameters by simultaneously fitting different sets of experimental signals using the least-squares regression method. Mathematically, this process involves minimizing the product of the root mean squared (RMS) differences between each set of experimental signals and their corresponding model predictions:

$$\psi(\boldsymbol{U}) = \prod_{j=1}^{M} \sqrt{\frac{1}{N_j} \sum_{i=1}^{N_j} \left( \frac{f_j(\boldsymbol{U}, \boldsymbol{P}, t_{d,i})}{E_j(t_{d,i})} - 1 \right)^2} = \prod_{j=1}^{M} \text{RMS}_j \quad (S16)$$

At the best fit, the gradient of $\psi$ should be zero for every element in $\boldsymbol{U}$:

$$\sum_{j=1}^{M} \left( \frac{\prod_{k \neq j} \text{RMS}_k}{2 \text{RMS}_j} \right) \left( \frac{1}{N_j} \sum_{i=1}^{N_j} \frac{2 \left( f_j(\boldsymbol{U}^*, \boldsymbol{P}^*, t_{d,i}) - E_j(t_{d,i}) \right)}{E_j(t_{d,i})^2} \frac{\partial f_j(\boldsymbol{U}^*, \boldsymbol{P}^*, t_{d,i})}{\partial u_l} \right) = 0,$$

$$\text{for } l = 1, 2, \ldots, n_u \quad (S17)$$

Here $\boldsymbol{P}^*$ is a random group of the possible control parameters since these input parameters have uncertainties, and $\boldsymbol{U}^*$ is the corresponding group of fitting parameters that make the best fit. The uncertainties of the unknown parameters can be revealed from the distribution of all the possible $\boldsymbol{U}^*$. Let us denote the mean values of all the possible $\boldsymbol{U}^*$ and $\boldsymbol{P}^*$ as $\boldsymbol{U}^0$ and $\boldsymbol{P}^0$, respectively. The function $f_j(\boldsymbol{U}^*, \boldsymbol{P}^*, t_{d,i})$ can be approximated by a first-order Taylor expansion around the point $(\boldsymbol{U}^0, \boldsymbol{P}^0)$ as:

$$f_j(\boldsymbol{U}^*, \boldsymbol{P}^*, t_{d,i}) \approx f_j(\boldsymbol{U}^0, \boldsymbol{P}^0, t_{d,i}) + \sum_{l=1}^{n_u} \frac{\partial f_j(\boldsymbol{U}^0, \boldsymbol{P}^0, t_{d,i})}{\partial u_l} (u_l^* - u_l^0)$$

$$+ \sum_{k=1}^{n_p} \frac{\partial f_j(\boldsymbol{U}^0, \boldsymbol{P}^0, t_{d,i})}{\partial p_k} (p_k^* - p_k^0), \text{ for } j = 1, 2, \ldots, M \quad (S18)$$

Substituting Eq. (S18) into Eq. (S17) and neglecting the higher-order terms, we can have:

$$\sum_{j=1}^{M} \left( \frac{\prod_{k \neq j} \text{RMS}_k}{\text{RMS}_j} \right) \left( \frac{1}{N_j} \sum_{i=1}^{N_j} \frac{1}{E_j(t_{d,i})^2} \left( f_j(\boldsymbol{U}^0, \boldsymbol{P}^0, t_{d,i}) - E_j(t_{d,i}) + \sum_{l=1}^{n_u} \frac{\partial f_j(\boldsymbol{U}^0, \boldsymbol{P}^0, t_{d,i})}{\partial u_l} (u_l^* - u_l^0) \right. \right.$$

$$\left. \left. + \sum_{k=1}^{n_p} \frac{\partial f_j(\boldsymbol{U}^0, \boldsymbol{P}^0, t_{d,i})}{\partial p_k} (p_k^* - p_k^0) \right) \left( \frac{\partial f_j(\boldsymbol{U}^0, \boldsymbol{P}^0, t_{d,i})}{\partial u_l} \right) \right) = 0,$$

$$\text{for } l = 1, 2, \ldots, n_u \quad (S19)$$

Eq. (S19) could be re-written in a matrix format as:

$$\sum_{j=1}^{M}\left(\frac{\prod_{k\neq j}\text{RMS}_k}{N_j\text{RMS}_j}\right)J_{U,j}^T G_j[F_j - E_j + J_{U,j}(U^* - U^0) + J_{P,j}(P^* - P^0)] = 0 \quad (S20)$$

where $G_j$ is the diagonal matrix($G_j = diag(\frac{1}{E_j(t_{d,1})^2}, \frac{1}{E_j(t_{d,2})^2}, \cdots, \frac{1}{E_j(t_{d,N_j})^2})$) and $E_j$ is the column vector of the $j$-th set of measured signals, and $F_j$ is the corresponding column vector of the signals evaluated by the thermal model at $(U^0, P^0)$, $J_{U,j}$ and $J_{P,j}$ are the Jacobian matrices of the function $F_j$ for variables $U$ and $P$, respectively:

$$J_{U,j} = \begin{pmatrix} \frac{\partial f_j(U^0, P^0, t_{d,1})}{\partial u_1} & \cdots & \frac{\partial f_j(U^0, P^0, t_{d,1})}{\partial u_{n_u}} \\ \vdots & \ddots & \vdots \\ \frac{\partial f_j(U^0, P^0, t_{d,N_j})}{\partial u_1} & \cdots & \frac{\partial f_j(U^0, P^0, t_{d,N_j})}{\partial u_{n_u}} \end{pmatrix} \quad (S21)$$

and

$$J_{P,j} = \begin{pmatrix} \frac{\partial f_j(U^0, P^0, t_{d,1})}{\partial p_1} & \cdots & \frac{\partial f_j(U^0, P^0, t_{d,1})}{\partial p_{n_p}} \\ \vdots & \ddots & \vdots \\ \frac{\partial f_j(U^0, P^0, t_{d,N_j})}{\partial p_1} & \cdots & \frac{\partial f_j(U^0, P^0, t_{d,N_j})}{\partial p_{n_p}} \end{pmatrix} \quad (S22)$$

Eq. (S20) could be rearranged as:

$$\sum_{j=1}^{M}\left(\frac{\prod_{k\neq j}\text{RMS}_k}{N_j\text{RMS}_j}\right)J_{U,j}^T G_j(E_j - F_j) - \sum_{j=1}^{M}\left(\frac{\prod_{k\neq j}\text{RMS}_k}{N_j\text{RMS}_j}\right)J_{U,j}^T G_j J_{P,j}(P^* - P^0)$$
$$= \sum_{j=1}^{M}\left(\frac{\prod_{k\neq j}\text{RMS}_k}{N_j\text{RMS}_j}\right)J_{U,j}^T G_j J_{U,j}(U^* - U^0) \quad (S23)$$

Let us denote

$$\Sigma_{UU} = \sum_{j=1}^{M}\left(\frac{\prod_{k\neq j}\text{RMS}_k}{N_j\text{RMS}_j}\right)J_{U,j}^T G_j J_{U,j} \quad (S23a)$$

$$\Sigma_{UP} = \sum_{j=1}^{M}\left(\frac{\prod_{k\neq j}\text{RMS}_k}{N_j\text{RMS}_j}\right)J_{U,j}^T G_j J_{P,j} \quad (S23b)$$

When $\Sigma_{UU}$ is non-singular, we can have $U^*$ explicitly expressed as:

$$U^* = \Sigma_{UU}^{-1}\sum_{j=1}^{M}\left(\frac{\prod_{k\neq j}\text{RMS}_k}{N_j\text{RMS}_j}\right)J_{U,j}^T G_j(E_j - F_j) - \Sigma_{UU}^{-1}\Sigma_{UP}(P^* - P^0) + U^0 \quad (S24)$$

The distributions of elements in $U^*$ can be obtained by calculating its covariance matrix. Since $E_j$ and $P^*$ are independent vectors, the covariance matrix of $U^*$ can be expressed as

$$\mathrm{Var}[\boldsymbol{U}^*] = \boldsymbol{\Sigma}_{UU}^{-1} \left[ \sum_{j=1}^{M} \left( \frac{\prod_{k \neq j} \mathrm{RMS}_k}{N_j \mathrm{RMS}_j} \right)^2 \boldsymbol{J}_{U,j}^T \boldsymbol{G}_j \mathrm{Var}[\boldsymbol{E}_j - \boldsymbol{F}_j] \boldsymbol{G}_j^T \boldsymbol{J}_{U,j} \right] \boldsymbol{\Sigma}_{UU}^{-1}$$
$$+ \boldsymbol{\Sigma}_{UU}^{-1} \boldsymbol{\Sigma}_{UP} \mathrm{Var}[\boldsymbol{P}^*] \boldsymbol{\Sigma}_{UP}^T \boldsymbol{\Sigma}_{UU}^{-1} \quad (S25)$$

Here $\mathrm{Var}[\boldsymbol{E}_j]$ is an $N_j$–by–$N_j$ diagonal matrix with the $i$-th component being $\left(E_j(t_{d,i}) - f_j(\boldsymbol{U}^0, \boldsymbol{P}^0, t_{d,i})\right)^2$ and $N_j$ being the total number of $t_d$ points, and $\mathrm{Var}[\boldsymbol{P}^*]$ is an $n_p$-by-$n_p$ diagonal matrix with the $k$-th component being $\sigma_{p_k}^2$ and $n_p$ being the total number of control parameters in $\boldsymbol{P}$.

Equation (S25) is the error propagation formula, which is a summation of two terms: the first term is the uncertainty from the experimental noise and fitting quality, and the second term is the uncertainty propagated from the errors of the control variables. The covariance matrix $\mathrm{Var}[\boldsymbol{U}^*]$ takes the format:

$$\mathrm{Var}[\boldsymbol{U}^*] = \begin{pmatrix} \sigma_{u_1}^2 & \mathrm{cov}[u_1, u_2] & \dots \\ \mathrm{cov}[u_2, u_1] & \sigma_{u_2}^2 & \dots \\ \vdots & \vdots & \ddots \end{pmatrix} \quad (S26)$$

where the elements on the principal diagonal $\sigma_{u_1}, \sigma_{u_2}, \dots, \sigma_{u_{n_u}}$ are the variance of the unknown parameters $u_1, u_2, \dots, u_{n_u}$; the off-diagonal ones $\mathrm{cov}[u_i, u_j]$ are the covariance of $u_i$ and $u_j$. If $\mathrm{cov}[u_i, u_j] = 0$, this means the variables $u_i$ and $u_j$ are entirely independent of each other.